# Four-Terminal Mechanically Stacked GaAs/Si Tandem Solar Cells

S. Hassan
University of Michigan

***Abstract*-This study investigates a four-terminal mechanically stacked double junction photovoltaic device based on GaAs as a top subcell and Si as a bottom subcell. Unlike two terminal monolithically series connected double junction photovoltaics, four-terminal mechanically stacked devices benefit from the ability to choose a combination of materials that are not constrained to lattice matching condition. GaAs top subcell is the best sensitive to visible light and Si bottom subcell is chosen to be grown on Si substrate which has relatively low cost. Moreover, the carriers generated by each subcell is collected independently to the external circuit. This electrical isolation of the subcells ensures higher efficiency, where no current matching nor tunnel junctions and related losses exist. A conversion efficiency of the device with a thickness in the order of 10 microns surpassed 27%.**

## I. INTRODUCTION

Nanodevices have been investigated for different electronic applications such as memory devices, sensors, thin film transistors, light emitting diodes and solar cells [1-8]. Solar cells enable the use of renewable, clean, and sustainable source of electrical energy. Multijunction solar cells have the potential to compete with fossil fuels due to their ultrahigh efficiency [9, 10]. But still they are not used commercially due to their high cost, which is directly related to the materials used in the cells. The choice of the materials used is constrained to the design and electrical configuration of the cell [11]. The conventional two-terminal series-connected monolithic multijunction is constrained to lattice and current matching conditions [11, 12]. Lattice matching condition hinders the choice of materials that best absorb the solar spectrum. Lattice mismatched materials cannot be used in the same cell, as this results in a high density of dislocations. Dislocations are active sites of non-radiative recombination, which limit the solar cell efficiency by reducing the carriers lifetime [13, 14]. Unlike the constrained two-terminal configuration, four-terminal device structure has a high degree of flexibility. Indeed, the choice of materials that best absorb the incident spectrum is broader, as there is no lattice matching requirement. Moreover, the carriers generated by each subcell is collected independently to the external circuit. This electrical isolation of the subcells ensures higher efficiency, where no current matching nor tunnel junctions and related losses exist. Thus four terminal mechanical stack was one of the configurations that boosted Shockley limit [15] of solar cell conversion efficiency [16].

This study investigates a four-terminal mechanically stacked double-junction photovoltaic device based on GaAs as a top subcell and Si as a bottom subcell. A lattice mismatch of about 4% exists between the materials used, but this lattice mismatch does not contribute to any losses in this novel device configuration.

## II. DEVICE DESIGN

Our choice of the material for the proposed device is GaAs (1.424 eV) for the top subcell and Si (1.124 eV) for the bottom subcell. This combination allows us to optimize the absorption of the solar spectrum in the visible region, GaAs is the most sensitive material to visible light, and Si is chosen to be grown on Si substrate which has relatively low cost. The GaAs is grown independently on GaAs substrate that is then etched and the GaAs active region is mechanically stacked in a four terminal configuration on the Si bottom subcell. The GaAs etched substrate can be reused in the growth of another subcell, which will decrease the cost of the final device manufacturing. A schematic of the proposed device is shown in Fig. 1.

The commercial simulation program PC1D [7] was used in the modeling and simulation of the device. The optimization of the device active layers for thicknesses and doping concentrations proceeded from the GaAs top subcell to the Si bottom subcell, i.e. from the top to the bottom of the device. The device optimization was performed under an illumination intensity of 0.1 W/cm$^2$, i.e. 1 sun, of the terrestrial AM1.5G spectrum. The device temperature was assumed to be constant at 25ºC, the simulated active area of the device was 1cm$^2$, and appropriate materials input parameters were used [7].

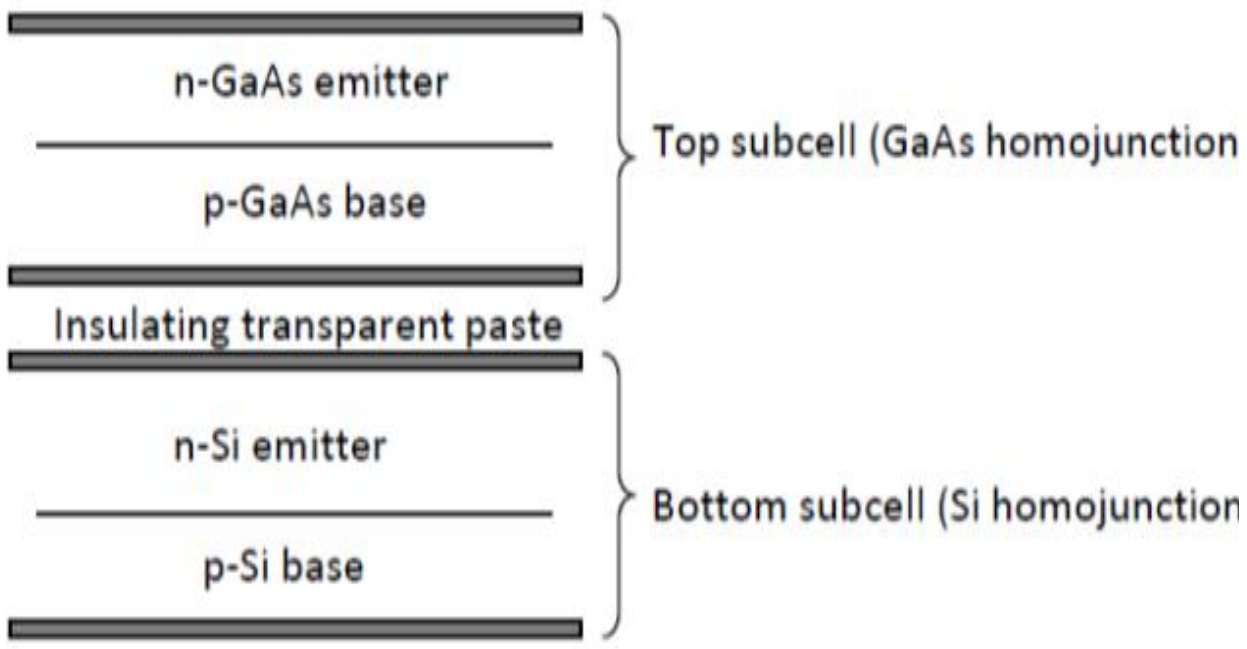

Figure 1: Schematic of a four-terminal mechanically stacked double-junction GaAs/Si.

## III. Device Optimization

The optimization of the thicknesses of the top subcell active layers was carried out taking into consideration the gain-loss balance between the top and bottom subcells. The optimal values obtained were for the whole tandem device and not individual subcells. A thickness of 0.1µm emitter layer of the GaAs top subcell is expected to be optimal. The gain-loss balance and the PV parameters i.e. efficiency, open circuit voltage, short circuit current density, and the fill factor are plotted in Fig. 2 versus the GaAs base layer. The tandem device efficiency (i.e. the top plus bottom subcells efficiencies) has its highest value of 27.55% at a base thickness between 0.1 and 0.2µm. The GaAs subcell can reach alone a value higher than 22% for a base thickness of 1µm, but this thickness results in an overall efficiency of 26.8% which is not optimal for the whole tandem. The thickness of the bottom subcell does not affect the top subcell, and thus a range of thicknesses between 5 and 20µm can be implemented giving a flexibility ragarding the active layers growth. A thickness of 10µm, i.e. 2µm for emitter and 8µm for base is chosen for the silicon cell to finalize the optimization and device operation.

The variation in the doping concentration of the top subcell active layers does not affect the bottom subcell. The PV parameters of the top subcell versus emitter and base doping concentrations are plotted in Fig. 3. The doping concentrations were varied between $10^{15}$ and $10^{18}cm^{-3}$, and the best performance was for a doping concentration of $10^{16}cm^{-3}$ predicted for both emitter and base of the top subcell. The doping concentrations of the bottom subcell active layers were also investigated. Optimal doping concentrations of $10^{18}$ and $10^{16}cm^{-3}$ were determined for the emitter and base of the bottom subcell respectively.

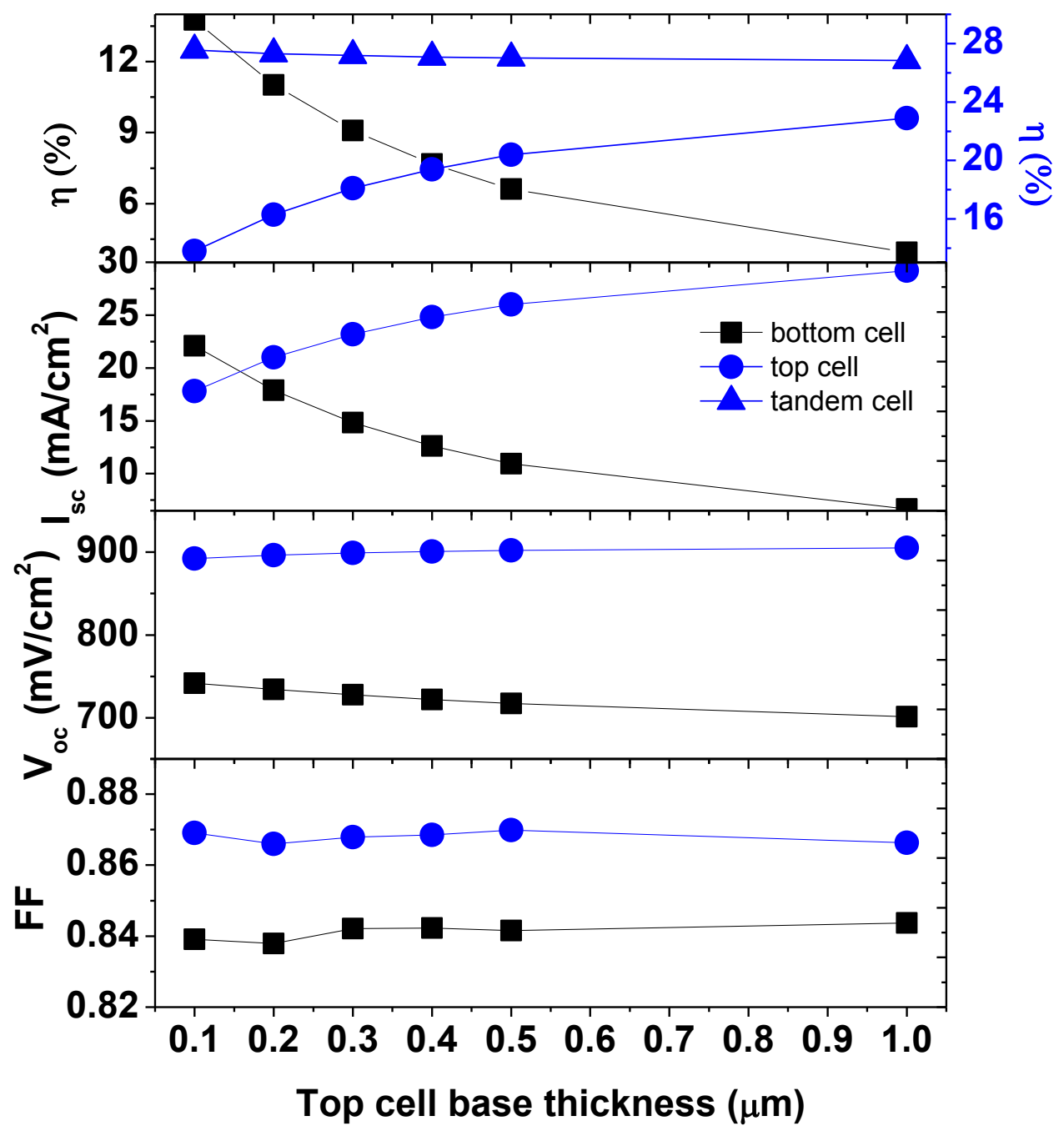

Figure 2: PV parameters as a function of the GaAs subcell base thickness.

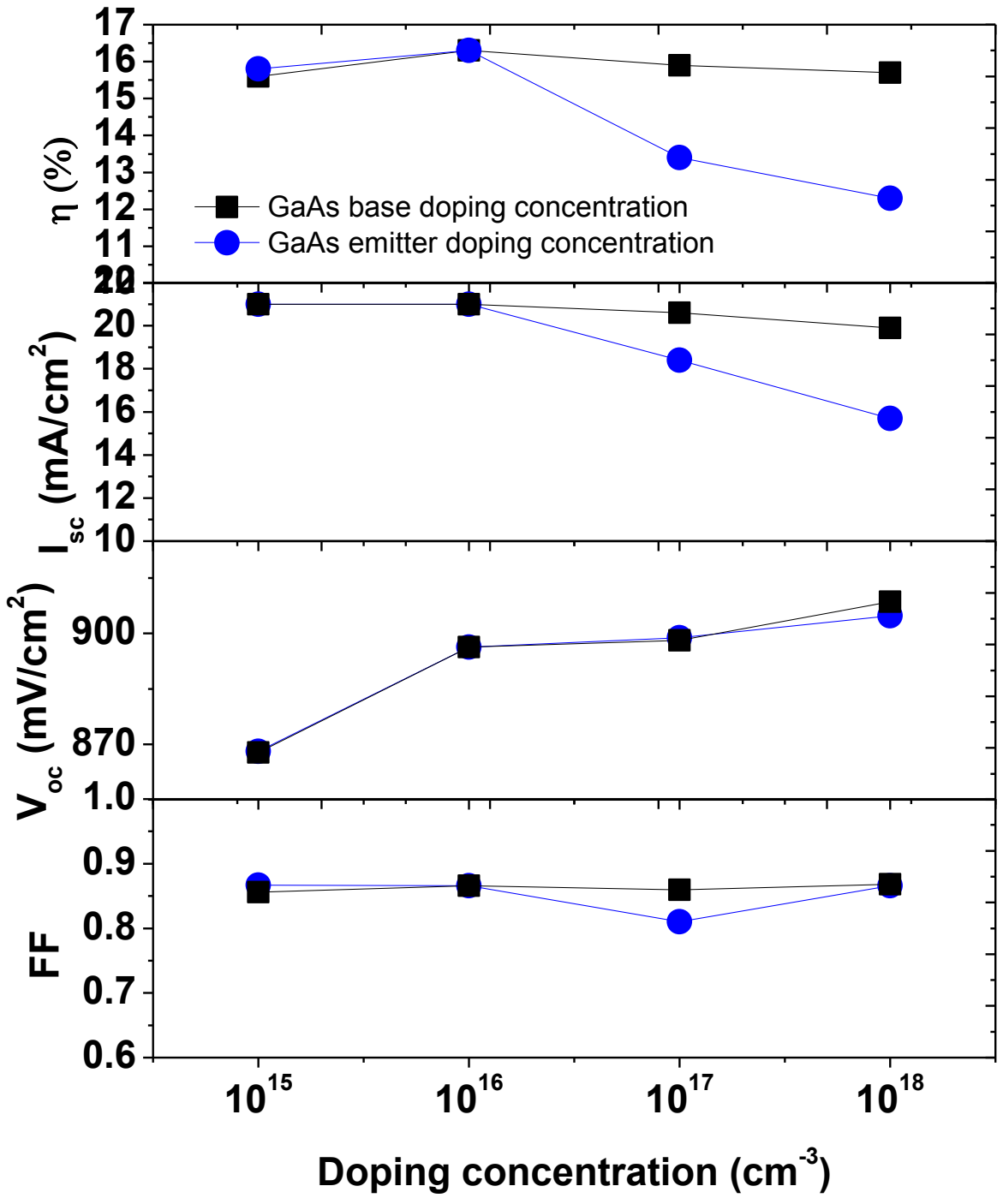

Figure 3: PV parameters of GaAs cell versus its emitter and base doping concentrations.

The I-V curves of the optimized device under one sun illumination and at 25ºC are plotted in Fig. 4. The current-voltage characteristics generated by the proposed device show the four-terminal configuration superiority, where the current-mismatch between top and bottom subcells is clear. The current densities will be much less if the same subcells are grown monolithically in a two terminal series connected configuration, due to the recombination that takes place at the resulting dislocation from the lattice-mismatch between both materials.

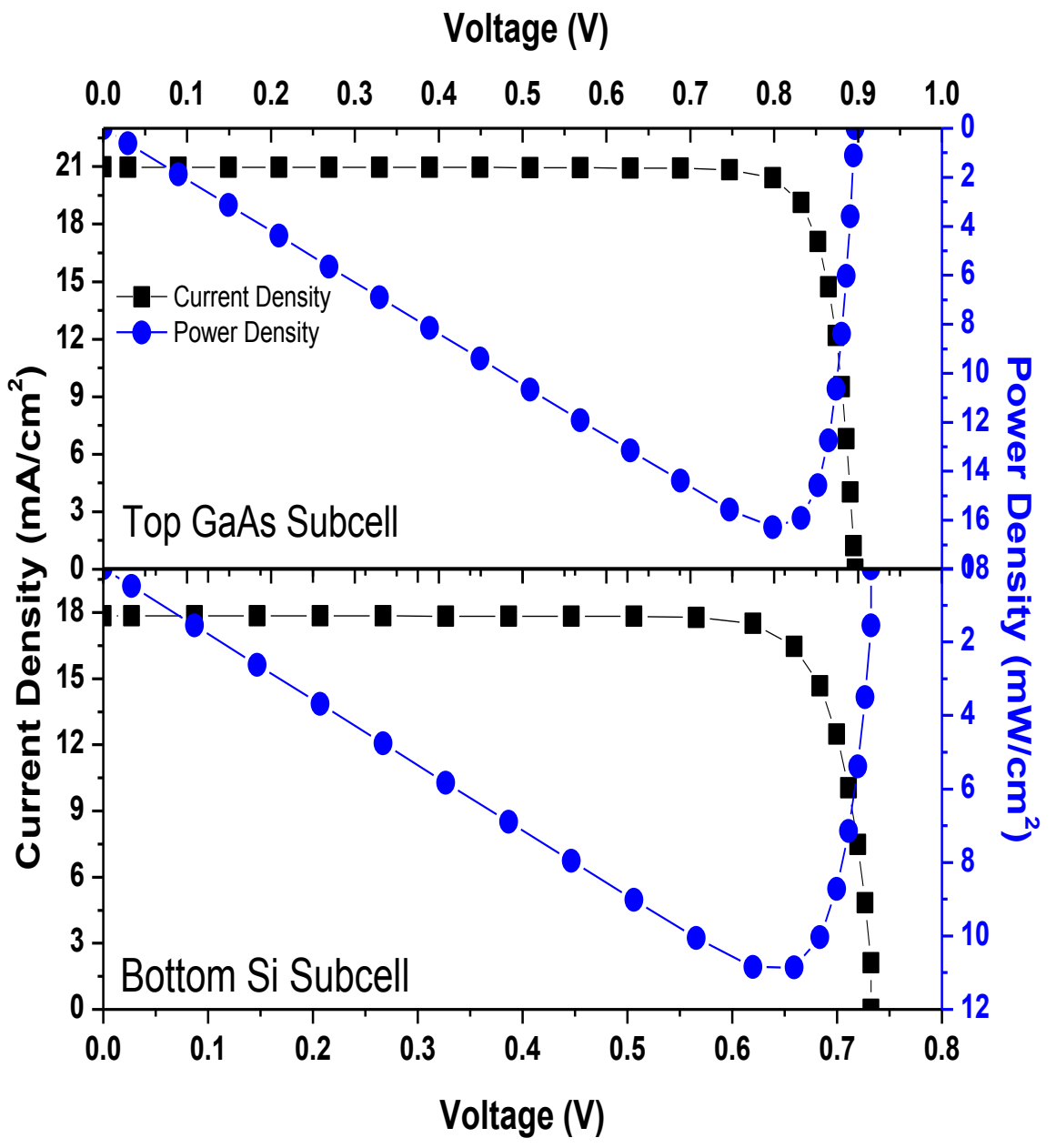

Figure 4: I-V curves of optimal top and bottom subcells.

## IV. CONCLUSION

A double-junction four-terminal mechanically stacked GaAs/Si photovoltaic device was simulated, and the potential of the device for CPV applications was investigated. The optimized device with a total thickness in the order of 10 microns is expected to reach a 27.55% conversion efficiency. The I-V curves of the optimized device were generated. The proposed device configuration benefit from the combination of appropriate materials without any constraints. The electrical isolation of the subcells through the four-terminal configuration allows the carriers to be independently collected without any losses that exist in the conventional two-terminal configuration. The proposed device was studied in a configuration that gives the highest flexibility possible for a multijunction solar cell design.

## REFERENCES

[1] Sleiman, A., M. F. Mabrook, R. R. Nejm, A. Ayesh, A. Al Ghaferi, M. C. Petty, and D. A. Zeze. "Organic bistable devices utilizing carbon nanotubes embedded in poly (methyl methacrylate)." *Journal of Applied Physics* 112, no. 2 (2012): 024509.

[2] Sleiman, A., M. C. Rosamond, M. Alba Martin, A. Ayesh, A. Al Ghaferi, A. J. Gallant, M. F. Mabrook, and D. A. Zeze. "Pentacene-based metal-insulator-semiconductor memory structures utilizing single walled carbon nanotubes as a nanofloating gate." *Applied Physics Letters* 100, no. 2 (2012): 023302.

[3] Sleiman, A., P. W. Sayers, and M. F. Mabrook. "Mechanism of resistive switching in Cu/AlOx/W nonvolatile memory structures." *Journal of Applied Physics* 113, no. 16 (2013): 164506.

[4] Nejm, Razan R., Ahmad I. Ayesh, Dagou A. Zeze, Adam Sleiman, Mohammed F. Mabrook, Amal Al-Ghaferi, and Mousa Hussein. "Electrical Characteristics of Hybrid-Organic Memory Devices Based on Au Nanoparticles." *Journal of Electronic Materials* (2015): 1-7.

[5] Sleiman, A., P. W. Sayers, D. A. Zeze, and M. F. Mabrook. "Two-terminal organic nonvolatile memory (ONVM) devices." *Handbook of Flexible Organic Electronics: Materials, Manufacturing and Applications* (2014): 413.

[6] Alaabdlqader, Homod S., Adam Sleiman, Paul Sayers, and Mohammed F. Mabrook. "Graphene oxide-based non-volatile organic field effect memory transistors." *IET Circuits, Devices & Systems* (2015).

[7] Sleiman, A., A. Albuquerque, S. J. Fakher, and M. F. Mabrook. "Gold nanoparticles as a floating gate in Pentacene/PVP based MIS memory devices." In *Nanotechnology (IEEE-NANO), 2012 12th IEEE Conference on*, pp. 1-5. IEEE, 2012.

[8] Sleiman, Adam Ahmad. "Two Terminal Organic Nonvolatile Memory Devices." PhD diss., Bangor University (Electronics), 2014.

[9] Emziane, M., and A. Sleiman. "Multi-junction solar cell designs." In *2011 IEEE GCC Conference and Exhibition (GCC).*

[10] Sleiman, Adam, and Mahieddine Emziane. "P/N/P Double-Junction GaAs/Ge Solar Cell Devices for PV and CPV." In *Sustainability in Energy and Buildings*, pp. 629-636. Springer Berlin Heidelberg, 2012.

[11] F. Dimroth and S. Kurtz, "High-efficiency multijunction solar cells," MRS Bull., vol. 32, pp. 230-235, 2007

[12] M. Yamaguchi, T. Takamoto, K. Araki, and N. Ekins-Daukes, "Multi-junction III-V solar cells: current status and future potentials," Sol. Energy, vol. 79, pp. 78-85, 2005.

[13] H. El Ghitani and S. Martinuzzi, " Influence of dislocations on electrical properties of large grained polycrystalline silicon cells. I. Model," J. Appl. Phys., vol. 66, pp. 1717, 1989.
[14] T. Kieliba, S. Riepe, and W. Warta, "Effect of dislocations on minority carrier diffusion length in practical silicon solar cells," J. Appl. Phys., vol. 100, pp. 063706, 2006.
[15] W. Shockley and H. J. Queisser, "Detailed Balance Limit of Efficiency of p-n Junction Solar Cells," J. Appl. Phys., vol. 32, pp. 510-520, 1961.
[16] L. Fraas *et al.*, "Over 35% efficient GaAs/GaSb stacked concentrator cell assemblies for terrestrial applications," *Conference record, 21st IEEE, PVSC,Kissimimee* pp.190-195, 1990.
[17] PC1D, version 5.9, School of Photovoltaic and Renewable Energy Engineering at the University of New South Wales, Australia.